\documentclass[pra,aps,showpacs,twocolumn]{revtex4}

\def\be{\begin{equation}}
\def\ee{\end{equation}}

\usepackage{graphicx}
\usepackage{bm}
\usepackage[T1]{fontenc}
\usepackage[latin9]{inputenc}
\usepackage{amsmath}
\usepackage{amssymb}
\usepackage{graphicx}
\usepackage{esint}

\begin{document}

\title{Enhanced trapping of colding lithium by using the multiple-sideband cooling in a two-dimensional magneto-optical trap}

\author{Kai Li$^1$, Dongfang Zhang$^1$, Tianyou Gao$^1$, Shi-Guo Peng$^1$}

\author{Kaijun Jiang$^{1,2}$}
\email{kjjiang@wipm.ac.cn}

\affiliation{$^1$State Key Laboratory of Magnetic Resonance and Atomic and
Molecular Physics, Wuhan Institute of Physics and Mathematics,
Chinese Academy of Sciences, Wuhan, 430071, China\\$^2$Center for
Cold Atom Physics,Chinese Academy of Sciences, Wuhan, 430071,
China}

\date{\today}

\begin{abstract}
 Trapping lithium with a big number in a simplified experimental setup has difficulties and challenges today. In this paper, we experimentally demonstrate the enhancement of \textsuperscript{6}Li trapping efficiency in a three-dimensional magneto-optical trap (3D MOT) by using the multiple-sideband cooling in a two-dimensional magneto-optical trap (2D MOT). To improve the number of trapped atoms, we broaden the cooling light spectrum to 102 MHz composed of seven frequency components and then trap atoms with a number of $6.0\times10^8$ which is about 4 times compared to that in the single-frequency cooling. The capture velocity and dependence of atomic number on the laser detuning have been analyzed, where the experimental result has a good agreement with the theoretical prediction based on a simple two-level model. We also analyze the loss rate of alkali metals due to fine-structure exchanging collisions and find that the multiple-sideband cooling is special valid for lithium.
\end{abstract}

\pacs{37.10.-x; 37.20.+j; 67.85.-d}
\maketitle

\section{Introduction}
Laser cooling and trapping \cite{Phillips1982PRL, Metcalf1999Newyork} has become the required first step to set up a cold atom laboratory on which people can carry on a diverge range of research, such as atomic interferometer \cite{Kaservich1991PRL}, high precision atomic clock \cite{Ye2014Nature, Ludlow2013Science}, precise spectroscopy \cite{Zelevinsky2015PRL}, quantum simulation \cite{Bloch2008RMP}, ultracold chemistry \cite{Jin2012ChemRev, Jin2011NaturePhysics}, and so on. In these applications, quantum degenerate Fermi gas has a prospective research potential due to its long-term lifetime in the strong interaction regime and tunability of the atomic interaction \cite{Chin2010RMP}, which affords us a reachable tabletop to study equation of state of strongly interacting Fermi gases \cite{Ueda2010Science, Salomon2010Nature}, fermionic polaron behavior \cite{Zwierlein2009PRL,Salomon2009PRL}, spin-orbit coupling effect \cite{Spielman2011Nature, Jiang2012PRA}, anisotropic characters of p-wave or higher partial wave interaction \cite{Esslinger2005PRL, Jiang2015PRL}, and other fundamental physics. Due to its big Feshbach-resonance width (about 300 GHz) \cite{Chin2010RMP, Grimm2005PRL}, $^6$Li is one of the most important Fermi spieces that can be extensively studied with quantum degeneracy in the experiment. With the aim to this observation, achieving a big number of trapped atoms is generally necessary \cite{Salomon2011EPJD}. Meanwhile with a high melting temperature (about 454 K), its vapor pressure is only significant at temperatures far above the maximum baking temperature of a conventional UHV vacuum system, and the trappable low velocity distribution is too small for us to obtain the satisfying number of trapped atoms. Generally, lithium is slowed using the Zeeman slow configuration \cite{Zhan1991Jpn, Hulet1992OL}, where fast-moving atoms are first slowed by resonant lasers with the help of magnet induced frequency shift and then loaded using a regular magneto-optical trap (MOT). But in this case, substantial engineering efforts are required to complete the system design and construction, where the oven has to be reloaded regularly due to the high flux of atomic beam in the setup. Also, lithium is chemically reactive with glass and thus the opposite glass window to the Zeeman slow needs special technical maintenance. So, how to build a simplified experimental setup to get a big number of trapped lithium atoms is still an open question and maintains challenging today.

In past decades, several research groups have tried different methods in improving lithium loading. Madison and colleagues collected lithium from a vapor chamber, where they can trap atoms with a number in the order of $10^7$ \cite{madison2013PRA}. In their experiment, an atomic oven was placed close to the trapping region to reduce loss due to transverse divergence, and an atomic block was set in the center line to avoid high-speed atoms kicking off the trapped atoms. Kasevich's group applied a comb-frequency laser to directly load lithium in a vapor chamber where the broadening spectrum (about 125 MHz) could increase the capture velocity and then couple faster atoms. They also got trapped atoms with a number in the $10^7$ order with this improved method \cite{Kasevich1994PRA}. When atoms are directly loaded from a room-temperature vapor cell, the atomic number is limited due to the small fraction of the low-velocity distribution. Walraven and colleagues applied a two-dimensional (2D) trap as the first cooling step and then pushed the slowed atoms to the three-dimensional (3D) trapping region \cite{Walraven2009PRA}. Using this innovated method, they achieved the atomic number similar to that in the Zeeman-slow configuration.

In this paper, we combine advantages both in the reference \cite{Kasevich1994PRA} and \cite{Walraven2009PRA}. We apply the multiple-sideband cooling with a spectrum of 102 MHz to slow fast moving atoms in a two-dimensional magneto-optical trap (2D MOT) first and then push the cooled atoms with a resonant laser beam to the 3D MOT. The atomic number in 3D is $6.0 \times 10^8$ and noticeably advanced (about 4 times) compared to that in the single-frequency cooling. We also get the dependence of atomic number on laser detuning, where a theoretical prediction based on a simple two-level model has a good agreement with the experimental results. The proposal demonstrated in this paper has following three advantages: 1), The multiple-sideband cooling can increase the trapped atomic number compared to that in the single-frequency cooling; 2), The 2D cooling stage can simplify the experimental setup compared to the Zeeman slow configuration; 3) This cooling method demonstrated here in lithium may be valid for other species which have difficulties to be directly loaded from a vapor chamber. Then we analyze the loss rate due to the fine-structure exchanging collision, explaining the reason for the success of the multiple-sideband cooling for lithium and suggesting the failure mechanism for earlier experiments on cesium.

The paper is arranged as follows. We first present experimental setup in Sec. II. Then we will show the experimental results and discussions in Sec. III. Finally, the main results are summarized in Sec. IV.

\section{Experimental Setup}
\begin{figure}
\centerline{\includegraphics[width=0.93\columnwidth]{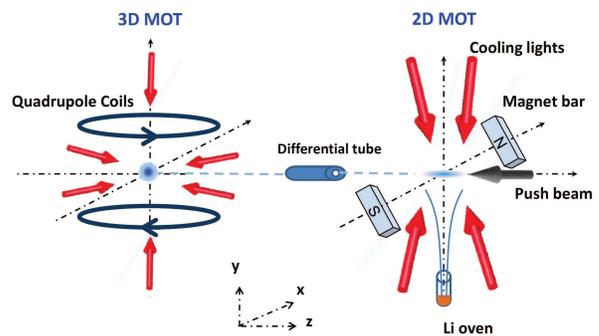}}
\caption{(color online) Schematics of the experimental setup. The whole system consists of two cooling and trapping regions, the 2D MOT and 3D MOT. The initial fast-moving atoms are emitted from a baked lithium oven which is located 170 mm below from the center of the 2D MOT. Atoms cooled in the 2D MOT are extracted toward the 3D MOT by using a pushing beam. The cooling light in the 2D MOT has multiple frequency sidebands, which has a spectrum width of 102 MHz to increase the capture velocity. Trapped atoms are detected in the 3D MOT using the fluorescence imaging method. The solid red arrows denote the cooling laser beams for 2D and 3D MOTs, and the black solid arrow means the push beam.}  \label{Fig1}
\end{figure}

The experimental setup is sketched in Fig.\ref{Fig1}. The main vacuum system consists of two stainless chambers for 2D and 3D MOTs, respectively. Due to small hyperfine splitters of the excited state ($2P_{3/2}$) of $^6$Li which are much smaller than the corresponding natural linewidth ($\Gamma=2\pi\times 5.87$ MHz), the cooling and repumping light should have comparable powers. Also considering that the hyperfine splitter of the ground state ($2S_{1/2}$) is small ($\Delta=2\pi\times 228$ MHz), these two light should be amplified separately because simultaneous amplification of two components with a small frequency difference in a taped amplifier (TA) will decrease the number of trapped atoms, which is implied in our previous work \cite{Jiang2013OL}. Under these considerations, the output of an external cavity diode laser (Topitca DL100@671nm 18 mW) is power amplified in a TA chip (Topitca BoosTA 670 L) and then equally divided into two beams. After passing through a series of acoustic optical modulators (AOMs), these two beams have a frequency difference of the hyperfine splitter of the ground state and then separately power amplified using other two TA chips, affording high powers for optical cooling and repumping, respectively. The laser frequency is locked on the atomic transition line using the Doppler-free saturated absorption spectroscopy in a lithium heat pipe. In the 2D MOT, the cooling and repumping light combine together in the x-y plane, cooling atoms in two dimensions. The line quadrupole magnetic field is supplied by two sets of $Nd_2Fe_{14}B$ magnet bars in the x axis, labeled with N and S, where the axial magnetic field is uniform and the radial magnetic gradient is modulated by changing the number of the magnet bars. A lithium oven (containing a mixture of 4 g $^6$Li enriched into 95\% and 2 g natural lithium with an abundance of 99\%) is connected to the main chamber using a stainless tube with a diameter of 16 mm and a length of 170 mm below the 2D MOT. Being heated to about 600 K, fast moving atoms are emitted from the oven upward, where the motions in x and y directions are slowed in the 2D MOT region and the velocity distribution in the z direction is narrowed during flying in the tube. A differential tube with a diameter of 3 mm and a length of 50 mm is placed between the two MOTs, which pumps the vacuum pressure gradient in 2 orders. The cooled atoms in the 2D MOT is pushed to the standard 3D MOT using a resonant beam with a power of about 1 mW, and then the trapped atoms in 3D are detected using the fluorescence imaging method.

\begin{figure}
\centerline{\includegraphics[width=0.93\columnwidth]{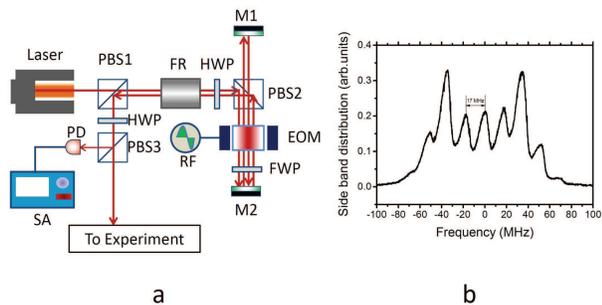}}
\caption{(color online) Generation of multiple sidebands. a, optical arrangement. PBS: polarization beam splitter, FR: Farady optical rotator, HWP: half-wave plate, FWP: four-wave plate, M1 and M2: mirrors, EOM: electro-optical modulator, RF: radio-frequency signal, PD: photodetector, SA: spectrum analyzer. b, power distribution of the multiple sidebands which is recorded with a SA.}  \label{Fig2}
\end{figure}

To improve the lithium loading, we broaden the spectrum of the cooling laser in the 2D MOT by generating multiple frequency sidebands with a small separation. We modulate the light phase in an EOM (Thorlabs EO-PM-NR-C1) as shown as in Fig.\ref{Fig2}(a). Considering that only the horizontally polarized light can be phase-modulated by a RF signal in the EOM, we make the light pass through the EOM four times in which for only two times the light has a horizontal linear polarization. Multiple sidebands are produced due to the nonlinearity effect in the EOM \cite{Kasevich1994PRA} and the output beam is spatially separated from the input. To uniquely determine the distribution of these sidebands, we use the heterodyne measurement to analyze the frequency-beat signal between the sidebands and an additional reference beam (not shown in Fig.\ref{Fig2}(a)) with a frequency separation of 300 MHz using a fast-responding photodiode (Newport 1621) \cite{Jiang2013OL,Kasevich1994PRA,Madison2007JOSAB}. The frequency beat signal is then recorded by a spectrum analyzer. Through choosing an optimal RF power, we can get six sidebands and a carrier frequency as shown in Fig.\ref{Fig2}(b), where the total power distributes comparably equally on the seven components. The spectrum bandwidth is 102 MHz under a RF driving signal with a frequency of 17 MHz. In fact, we also modulate the repumping light with a 12 MHz RF signal and get a spectrum bandwidth of 72 MHz. In the following discussion, we only include the sideband effect of the cooling light for simplification.

\section{Experimental results and discussion}

\begin{figure}
\centerline{\includegraphics[width=0.93\columnwidth]{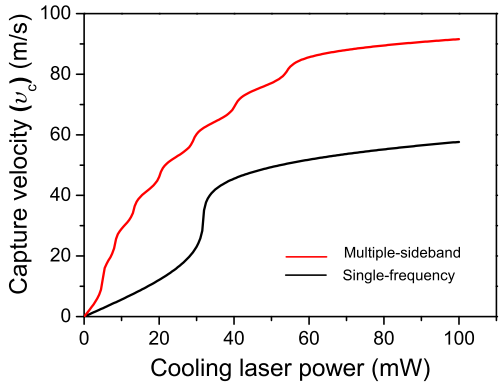}}
\caption{(color online) Capture velocity versus the total cooling light power in the 2D MOT. The red line indicates the capture velocity in the multiple-sideband cooling while the black line means the single-frequency cooling.}  \label{Fig3}
\end{figure}

\begin{figure}
\centerline{\includegraphics[width=0.93\columnwidth]{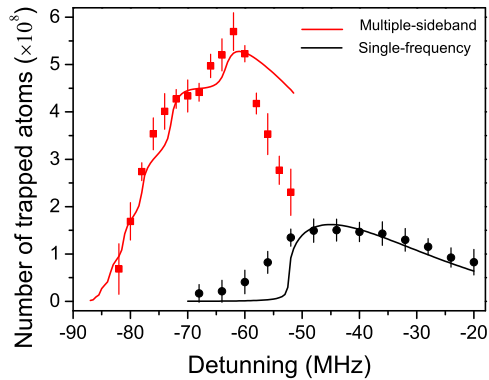}}
\caption{(color online) Number of trapped atoms in the 3D MOT. The red squares and the black circles denote the multiple-sideband cooling and single frequency cooling, respectively. Each experimental point comes from five measurements and the error bar is the standard deviation (SD). The solid red and black lines are the theoretical predictions from Eq.(\ref{forcemultiple}) and Eq.(\ref{forcesingle}), respectively.}  \label{Fig4}
\end{figure}

In a vapor-cell trap, atoms with velocities less than the capture velocity $v_c$ are slowed sufficiently after entering the intersecting laser beams thus they can be loaded into a MOT. So the capture velocity $v_c$ plays a very important role in determining the atom trapping efficiency. Solving the steady-state rate equation $\partial N/\partial t = R-\Gamma_c N=0$, we get the number of trapped atoms \cite{Wieman1992PRA, Steven1992OL}:

\begin{equation}
   N = \frac{R}{\Gamma_c} = 0.1 \frac{A}{\sigma} \left( \frac{v_c}{v_{thermal}} \right)^4, \label{number}
\end{equation}

\noindent where $R$ is the capture rate, $\Gamma_c$ is the collision rate between trapped atoms with atoms in the background gas,  $\sigma = 4.4\times 10^{-14}$ cm$^2$ \cite{Walraven2009PRA} is the total cross section for these collisions, $A=\pi d^2$ is the trap surface where $d$ is the diameter of the laser beam, and $v_{thermal} = 1418.8$ m/s is the average velocity of the background gas with a temperature of 570.5 K. This calculation neglects the contribution of intratrap collisions to the loss rate of the trap. For the densities of background vapor and trapped atoms at which we operate, this contribution is relatively small and thus does not affect most of the comparisons we make between our theoretical model and experimental data. In calculating the capture velocity $v_c$, we only compute the one-dimensional slowing force on atoms in the trapping region because this force is nearly the same along three axes. For simplification, we treat the atom as a two-level system and ignore effects of the magnetic field and the Gaussian intensity profile of laser beams. Then we can write the radiation-pressure force for the single-frequency cooling,

\begin{eqnarray}
   F = \frac{\hbar k \Gamma}{2}\left[\frac{I/I_s}{1+I/I_s+\left(\frac{2(\Delta_0-kv)}{\Gamma}\right)^2} \right.\nonumber \\
    \left. - \frac{I/I_s}{1+I/I_s+\left(\frac{2(\Delta_0+kv)}{\Gamma}\right)^2} \right], \label{forcesingle}
\end{eqnarray}

\noindent and for the multiple-sidebands cooling,

\begin{eqnarray}
  F = \frac{\hbar k \Gamma}{2}\sum_{i=1}^7 \left[\frac{I/I_s}{1+I/I_s+\left(\frac{2(\Delta_i-kv)}{\Gamma}\right)^2} \right. \nonumber \\
    \left. - \frac{I/I_s}{1+I/I_s+\left(\frac{2(\Delta_i+kv)}{\Gamma}\right)^2} \right], \label{forcemultiple}
\end{eqnarray}

\noindent where $\hbar$ is Plank's constant $h$ divided by $2\pi$, $k = 9.36\times 10^4$ cm$^{-1}$ is the wave number of the trapping light, $\Gamma = 2\pi \times 5.87$ MHz is the linewidth of the excited state, $I$ is the intensity of each beam and $I_s = 2.54$  mW/cm$^2$ is the saturation intensity. $\Delta_0$ is the laser detunning to the resonant transition in the single frequency cooling. In the multiple-sidebands cooling, $\Delta_4$ is the laser detunning of the carrier frequency, and $\Delta_i = \Delta_4 + (i-4)\times q$ where $q = 17$ MHz is the frequency separation between two adjacent components. There are total seven components, which means that $i = 1, 2, \cdots, 7$. An atom with an initial velocity $v_c$ will be captured in the 3D MOT with the help of the pushing beam if, during its flight through the 2D MOT cooling beams, its velocity is reduced to $v = 0.0$ m/s by the scattering force. By numerically solving the Eq.(\ref{forcesingle}) and Eq.(\ref{forcemultiple}), we can calculate the capture velocities as a function of the total cooling light power, which is shown in Fig.\ref{Fig3}. In the multiple-sideband cooling, we assume that the total power distributes equally into the seven components for simplification. The diameter of the laser beam is 4 cm. $\Delta_0 = 45$ MHz and $\Delta_4 = 65$ MHz are the single-frequency cooling and the multiple-sideband cooling detunning, respectively, which are close to the optimal values as shown in Fig.\ref{Fig4}. It is obvious that the capture velocity in the multiple-sidebands cooling is dramatically increased compared to that in the single-frequency cooling. For example when the cooling light power is 60.0 mW, the capture velocity in the multiple-frequency cooling is 85.6 m/s  while it is only 51.8 m/s in the single-frequency cooling.

Compared to the common cooling method, the characteristic advantage of the multiple-sideband cooling is that it can slow faster atoms and then increase the atom trapping efficiency. In laser cooling, it is required to keep the cooling laser frequency resonant with moving atoms including the Doppler shift. So exploring the dependence of the atom trapping efficiency on the laser detunning ($\Delta_0$ or $\Delta_4$) can demonstrate the superiority of the multiple-sideband cooling. In the experiment, it is very difficult to directly measure the capture velocity, while probing the number of trapped atoms is easy. We probe the number of trapped atoms in the 3D MOT which are pushed away from the 2D MOT. To compare the atom trapping efficiencies in the single frequency and multiple-sideband cooling, we fix all the experimental parameters same except frequency components in the cooling light. In the 2D MOT, the diameter of the laser beam is 4 cm, the total cooling power is 40 mW, and the magnetic field gradient in the radial direction is 50 G/cm which is similar to that in the reference \cite{Walraven2009PRA}; In the 3D MOT, the diameter of the laser beam is 3 cm, the total cooling power is 50 mW, and the magnetic field gradient in the radial direction is 7 G/cm. We just scan the laser detunning with the help of a AOM in the double-pass configuration and detect the number of trapped number using the fluorescence imaging method with a digital CCD during a 1 ms exposure. The results are shown in Fig.\ref{Fig4}. In the multiple-side bands cooling, the theoretical prediction has been multiplied with 2.5 to match the experimental data, while the normalization factor in the single frequency cooling is 4.6 which is close to 3.3 in reference \cite{Wieman1992PRA}. The underestimation of the theoretical prediction mainly comes from the effect of the magnetic field which may increase the atomic trapping efficiency \cite{Wieman1992PRA}. The simple model based on a two-level system qualitatively agrees with the experimental data, predicting the main trend of the variation of the atomic number versus the laser detunning. The optimal detunning (about 61 MHz) in getting the maximal atomic number in the multiple-sideband cooling is much bigger than that in the single frequency cooling (about 44 MHz), which indicates that faster atoms can be slowed in our proposed scheme. Also, the maximum number of trapped atoms has been obviously advanced.

In the multiple-sideband cooling, both theoretical prediction and experimental result show a step-like increasing behavior when the laser detunning being reduced toward the optimal value, which is also shown in reference \cite{Kasevich1994PRA} and Fig.\ref{Fig3}. This step-like behavior comes from the discrete distribution of the seven frequency components with a 17 MHz separation. Our calculation further shows that the step-like phenomenon will become invisible when decreasing the frequency separation. In addition, when the frequency detunning ($\Delta_4$) is close to 51 MHz and thus the sideband $\Delta_7$ simultaneously become resonant to the atomic transition, the loss rate due to light-assisted collisions will increase, which isn't included in our theoretical calculation. This might explain the reason that the experimental results decrease versus the laser detunning more rapidly than the theoretical prediction in the small-detunning region for the multiple-sideband cooling. From our calculation, the capture velocity can even become negative for a $\Delta_4$ less than 51 MHz, which means that the photon scattering force will accelerate atoms and Eq.(\ref{number}) will no longer be valid to predict the number of trapped atoms.

\begin{figure}
\centerline{\includegraphics[width=0.93\columnwidth]{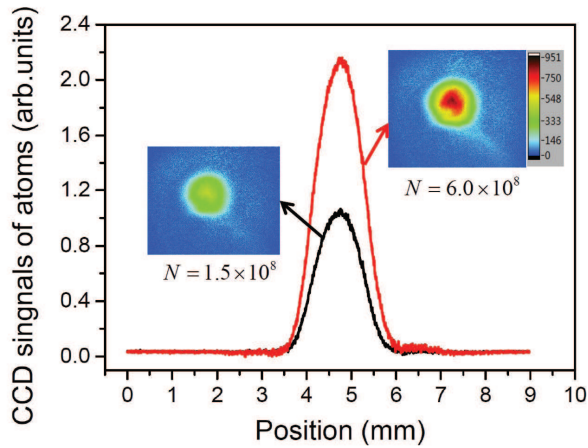}}
\caption{(color online) Fluorescence images of  trapped atoms in the 3D MOT. The red and black solid lines are the integrated fluorescence along the vertical direction for the multiple-sideband and single-frequency cooling, respectively. Corresponding fluorescence images are also inserted on the plot, as well as the number of trapped number: $N = 6.0\times 10^8$ for the multiple-sideband cooling and $N = 1.5\times 10^8$ for the single-frequency cooling. The color in CCD images is scaled to the fluorescence amplitude where the saturation value is $2^{12} = 4096$ and the background electric noise is less than 200.}  \label{Fig5}
\end{figure}

In order to quantitatively demonstrate the enhancement of the atom trapping efficiency, we optimize the magnetic field gradient and the laser detunning in the 2D MOT to get the optimal number of trapped atoms, while other parameters are kept the same with Fig.\ref{Fig4}. The trapped atoms are shown in Fig.\ref{Fig5} and the corresponding experimental parameters are shown in Table \ref{parameter}. The atomic number in the multiple-sideband cooling is $6.0\times 10^8$ which is 4.0 times compared to that in the single frequency cooling, while in reference \cite{Kasevich1994PRA} the advance factor is 3.1. The center and size of trapped atoms in these two kinds of cooling are almost the same within $\pm30$ $\mu$m, which is mainly resulted from the same parameters in the 3D MOT \cite{Madison2007JOSAB}. The atomic number in Fig.\ref{Fig5} is from one measurement, while in Table \ref{parameter} it is the averaging result within five measurements.

\begin{table}
  \centering
  \vspace{-1mm}
  \caption{Optimal parameters in the 2D MOT and the maximal atomic number in 3D. $q$ is the frequency separation, and No is the sideband number.}\label{parameter}
\begin{tabular}{l|cc}
\hline \hline parameters & single-frequency & multiple-sideband  \\ \hline
cooling detunning & 44.0 MHz & 63.8 MHz \\
cooling sideband $q$ & & 17.0 MHz \\
cooling sideband No & & 6 \\
repumping detunning & 14.1 MHz& 40.2 MHz  \\
repumping sideband $q$ & & 12.0 MHz \\
repumping sideband No & & 6 \\
magnetic gradient & 50.0 G/cm & 45.0 G/cm \\
atomic number in 3D & 1.5(0.5)$\times 10^8$ & 6.1(0.9)$\times 10^8$ \\ \hline \hline
\end{tabular}
\end{table}

The number of atoms loaded into the trap after a fixed time depends not only on the loading rate, but also on the rate at which atoms are ejected from the trap. In general, the presence of nearly resonant, red-detuned light produces fine-structure exchanging collisions by inducing transitions to excited molecular states \cite{Julienne1992PRA}. So the multiple-sideband cooling can enhance the capture velocity on one side, but on the other side it will increase the loss rate due to fine-structure exchanging collisions. Assuming that atoms are nearly at rest before collisions, we calculate the resulting velocity due to the fine-structure exchanging collision, $2\pi\times \hbar \Delta_{FS} = 2mv^2_{FS}/2$, where $\Delta_{FS}$ is the fine-structure splitting. The results are shown in Table \ref{loss} for different alkali atoms. For each kind of isotope, we only calculate one kind of atom because $v_{FS}$ of each one in the same isotope is almost the same because of the similar fine-structure splitting and atomic mass. Due to its small fine-structure splitting (about 10 GHz), $v_{FS}$ of lithium is much smaller than the capture velocity under a general experimental condition as shown in Fig.\ref{Fig2}. So lithium after the fine-structure exchanging collision can be recaptured in the multiple-sideband cooling. For sodium, $v_{FS}$ is 94.9 m/s and the capture velocity in the multiple-sideband cooling can reach this value with a possibility. In fact, M. Zhu and colleagues have cooled sodium using the broadband-laser cooling which is similar to the multiple-sideband cooling \cite{Hall1991PRL}. For potassium, rubidium or cesium, $v_{FS}$ is quite big, far beyond the general experimental conditions. This may explain the failure of earlier efforts to trap cesium with the broadband-laser cooling or multiple-sideband cooling \cite{Wieman1992PRA, Steven1992OL}. But for these heavier metals, multiple-sideband cooling might be available if the central area of the laser beam is blocked, thus producing a region where trapped atoms can accumulate without suffering additional sideband induced losses \cite{Kasevich1994PRA, madison2009JOSA,madison2013PRA}.

\begin{table}
  \centering
  \vspace{-1mm}
  \caption{Resulting velocity due to fine-structure exchanging collisions for different alkali atoms.}\label{loss}
\begin{tabular}{r|ccccc}
\hline \hline  & $^6$Li & $^{23}$Na & $^{40}$K & $^{87}$Rb & $^{133}$Cs \\ \hline
$\Delta_{FS}$ (THz) & 0.0101 & 0.5162 & 1.7301 & 7.1230 & 16.6097 \\
$v_{FS}$ (m/s) & 26.0 & 94.9 & 131.7 & 181.2 & 223.8 \\ \hline \hline
\end{tabular}
\end{table}

\section{Conclusion}
In conclusion, we experimentally demonstrate the enhancement of lithium trapping efficiency in the 3D MOT by using the multiple-sideband cooling in the 2D MOT. The trapped atomic number is about 4 times compared to that in the single-frequency cooling. The dependence of the atomic number on the cooling light detuning has been analyzed, where the experimental result has a good agreement with a theoretical prediction based on a simple two-level model. We also find that the multiple-sideband cooling is special valid for lithium due to its small fine-structure splitting and propose that this cooling scheme might become available for other alkali metals using an atomic block.

\begin{acknowledgments}
This work is supported by NSFC (Grants No. 11434015, No. 91336106, No. 11204355, No. 11474315, and No. 11004224), NBRP-China (Grant No. 2011CB921601), and programs in Hubei province (Grants No. 2013CFA056).
\end{acknowledgments}

\end{document}